\documentclass{elsart}
\usepackage{amsfonts}

\usepackage{graphicx,amssymb,amsmath}
\usepackage[a4paper]{geometry}
\usepackage{ulem}
\usepackage{color}
\usepackage{rotating}

\begin{document}

\renewcommand{\v}[1]{{\bf #1}} \renewcommand{\c}[1]{{\cal #1}} %
\renewcommand{\b}[1]{{\bar{ #1}}}

\renewcommand{\>}{\rangle} \renewcommand{\Im}{{\rm Im}} \renewcommand{\Re}{{%
\rm Re}} \renewcommand{\t}[1]{{\tilde #1}} %\newcommand{\e}{\epsilon}
\renewcommand{\th}{\theta}
\newcommand {\C}{\textcolor {red}}

\begin{frontmatter}

\title{Topological term in the non-linear $\sigma$ model of the SO(5) spin chains}

\author{Dung-Hai Lee}
\address{
Department of Physics, University of California at Berkeley,
Berkeley, CA 94720, USA;}
\address{Materials Sciences Division, Lawrence Berkeley National Laboratory, Berkeley, CA 94720, USA}
\author{Guang-Ming Zhang}
\address{Department of Physics, Tsinghua University, Beijing 100084,
China}
\author{Tao Xiang}
\address{Institute of Physics and Institute of Theoretical Physics,
Chinese Academy of Sciences, Beijing 100190, China.}

\begin{abstract}
We show that there is a topological (Berry phase) term in the
non-linear $\sigma$ model description of the SO(5) spin chain. It
distinguishes the linear and projective representations of the
SO(5) symmetry group, in exact analogy to the well-known
$\theta$-term of the SO(3) spin chain. The presence of the
topological term is due to the fact that $\pi_2(
\frac{SO(5)}{SO(3)\times SO(2)})= \mathbb{Z}$. We discuss the
implication of our results on the spectra of the SO(5) spin chain,
and connect it with a recent solvable SO(5) spin model which
exhibits valence bond solid ground state and edge degeneracy.
\end{abstract}

%\date{\today}


\end{frontmatter}
%\pacs{03.65.Vf, 05.30.-d, 75.10.Pq} \maketitle

\newpage

\section{Introduction}

The effects of topological terms on the dynamics of Goldstone modes and the
quantum number of solitons and instantons in non-linear ${\sigma }$ (NL${%
\sigma }$) models have a long history, and continue to attract strong
interests from the physics community\cite{wilczek}. For example, in one
spatial dimension (1D), quantum SO(3) spin chains have fundamentally
different low energy properties, depending on whether the site
representation is linear (spin integer) or projective (spin half-odd
integer). For nearest neighbor isotropic exchange interaction, the former
cases always have excitation gaps while the latter are gapless - the well
known Haldane conjecture\cite{haldane}. Aside from the usual stiffness
terms, the (1+1)D NL${\sigma }$ models for these spin chains contain a
topological (Berry phase) term (also known as the $\theta $ term)\cite%
{haldane}. When the space-time configuration of the Neel order parameter
wraps the target space\cite{mermin} ($S^{2}$) $~n$ times, the Berry phase factor is +1
for integer spin chains, while it is $(-1)^{n}$ for half-odd integer spin
chains\cite{haldane}.

Using an algebraic approach, Chen, Gu and Wen\cite{xc} recently generalized
an idea of Ref.\cite{pollmann} and argued that, in one spatial dimension, a
gapped ground state which is invariant under translation and the global
symmetry operation (we refer to this type of state as ``totally symmetric''
in the following) is obtainable when the site representation of the global
symmetry group (which can be discrete) is linear. When the site
representation is projective, a totally
symmetric ground state must be gapless. In a projective representation $D$,
the matrix product $D(g_{1})D(g_{2})$, where $g_{1,2}$ are group elements,
can differ from $D(g_{1}.g_{2})$ by a phase factor $e^{i\theta
(g_{1},g_{2})} $. For the SO(3) group, an integer spin forms the linear
representation while a half-odd integer spin forms the projective
representation. Hence the spectral difference between the translational
invariant integer and half-odd integer SO(3) spin chains constitutes a
special example of the results of Ref.\cite{xc}. Thus there are two ways to
view the difference between integer and half-odd integer SO(3) spin chain:
one is geometrical (the Berry's phase)\cite{haldane} and the other is
algebraic\cite{xc}.

SO(5) is a rank-2 classical Lie group. It also has linear and
projective representations. For example, in the vector
representation, the generators of the Lie algebra are given
by\cite{georgi}
\begin{equation}
(L_{ab})_{jk}=-i\delta _{a,j}\delta _{b,k}+i\delta _{a,k}\delta _{b,j},
\label{vec}
\end{equation}%
where $a,b,=1,..5$, and $i,j=1..5$. Two consecutive $\pi $ rotations
generated by, e.g., $L_{12}$ give
\begin{equation}
U_{12}(\pi )U_{12}(\pi )=I_{5\times 5}.
\end{equation}%
The spinor representation, on the other hand, is given by\cite{georgi}
\begin{equation}
L_{ab}=i[\Gamma _{a},\Gamma _{b}]/4,  \label{sp}
\end{equation}%
where $\Gamma _{a,b}$ are the $4\times 4$ gamma matrices (e.g., $\Gamma
_{1}=-\sigma _{y}\otimes \sigma _{x}$, $\Gamma _{2}=-\sigma _{y}\otimes
\sigma _{y}$, $\Gamma _{3}=-\sigma _{y}\otimes \sigma _{z}$, $\Gamma
_{4}=\sigma _{x}\otimes I_{2\times 2}$, $\Gamma _{5}=\sigma _{z}\otimes
I_{2\times 2}$). In this case it is simple to check that two consecutive $%
\pi $ rotations generated by $L_{12}$ yield
\begin{equation}
U_{12}(\pi )U_{12}(\pi )=-I_{4\times 4}.
\end{equation}%
Thus the vector representation is linear, while the spinor
representation is projective. According to Ref.\cite{xc}, a spin
chain of the former type can have a totally symmetric ground state
with a gapped spectrum, while a spin chain of the latter type has
to be gapless if it is totally symmetric. In the following, we
will seek for the geometric (Berry' phase) difference between the
two cases.

Another motivation for us to study the Berry's phase of the SO(5) spin chain
is a recent exactly solvable 1D SO(5) spin model (in the vector
representation) proposed by Tu et. al.\cite{zx}. The ground state is a
translational invariant matrix product state, i.e., a valence bond solid
state\cite{aklt}, and the excitation spectrum has a gap. Moreover, when the
chain is subjected to open boundary condition, there are edge states.
These properties are reminiscent of those of integer SO(3) spin chains%
\cite{ng}.

With the advance of cold atom physics, the SO(5) spin chain might
not be a purely academic model anymore. An SO(5) symmetric spin chain can in
principle be realized experimentally when the hyperfine spin-3/2 cold
fermions on an 1D optical lattice form the Mott-insulating state\cite{tu}.
At quarter filling (one fermion per site), the effective spin chain is in
the spinor representation, while for half-filling (two fermions per
site) it is in the SO(5) vector representation.
Therefore the idea presented here might one day be tested experimentally.

\section{Model formulation}

Let us start by considering the following SO(5) invariant Hamiltonian
\begin{equation}
H=\sum_{i}\left[ J_{1}\left( \sum_{a<b}L_{ab}^{i}L_{ab}^{i+1}\right)
+J_{2}\left( \sum_{a<b}L_{ab}^{i}L_{ab}^{i+1}\right) ^{2}\right]   \label{h5}
\end{equation}%
where $L_{ab}$'s are the SO(5) generators and $J_{1,2}>0$. When $L_{ab}$ are
given by Eq.(\ref{vec}), the ground state is translational invariant and the
spectrum has a gap in the parameter range $1/9<J_{2}/J_{1}<1/3$\cite{zx}.
Naively, one would not expect the NL$\sigma $ model action of this model to
contain a topological term. This is because in contrast to SO(3) spin chain
where the space-time dimension ($1+1$) matches the dimension of the target
space of the order parameter ($S^{2}$), for the SO(5) spin
chain the  target space dimension is much larger than the
space-time dimension.

To understand the structure of the target space for the SO(5) spin
chain, we need to know how the presence of SO(5) ``magnetic''
moment breaks the global SO(5) symmetry. For that purpose, it is
sufficient to consider the following mean-field theory where
non-linear terms in $L_{ab}$ are decoupled into linear ones with
order parameter given by $\langle L_{ab}^{i}\rangle
=(-1)^{i}m_{ab}$
\begin{equation}
H_{\text{MF}}=\left( -2J_{1}+2J_{2}\Delta ^{2}\right) \sum_{i,a<b}(-1)^{i}m_{%
\text{ab}}L_{\text{ab}}^{i}+\sum_{i}\left( J_{1}\Delta ^{2}-3J_{2}\Delta
^{4}\right) ,
\end{equation}%
where $\Delta ^{2}=\sum_{a<b}m_{ab}^{2}$. The question at hand is,
for a fixed total magnitude of $m_{ab}$ (i.e. fixed
$\sum_{a<b}m_{ab}^{2}$), what is the most energetically favorable
ratio between different components of $m_{ab}$. This can be
answered by diagonalizing a single-site Hamiltonian
\begin{equation}
H_{1}=-\sum_{a<b}m_{ab}L_{ab}.  \label{eq:h1}
\end{equation}%
and see what ratio gives the lowest ground state energy. (Of course we need
to remember the sign of $m_{ab}$ change from site to site.)

\begin{figure}[tbp]
\begin{center}
\includegraphics[angle=90,scale=0.5]{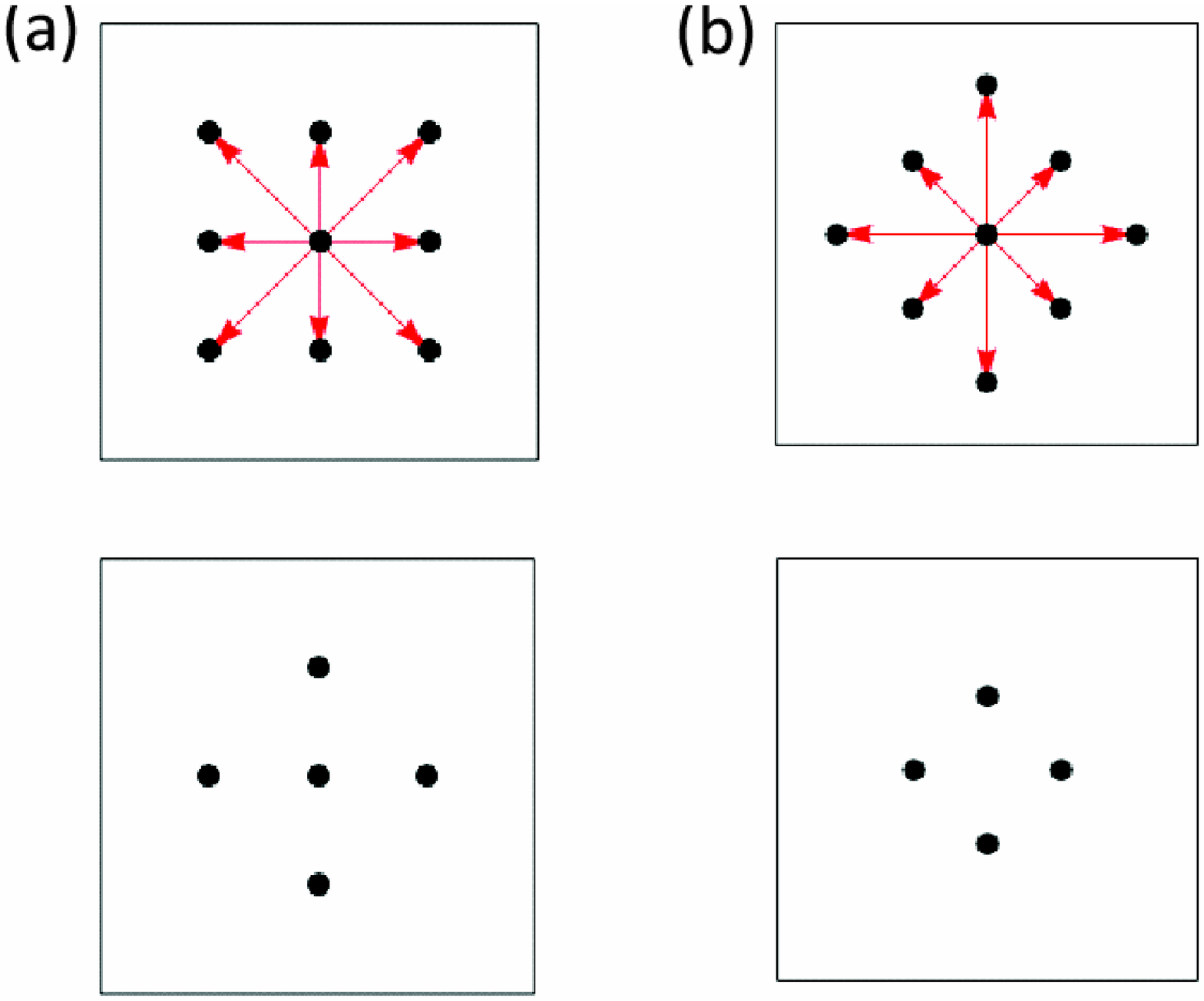}
\end{center}
\caption{(color on-line) The root (upper row) and weight (lower row)
patterns of the vector (a) and spinor (b) representation of SO(5). The x and
y coordinates of each dot corresponds to the eigenvalue of $H_{1}$ and $H_{2}
$. The red arrows indicate how the rasing/lowering operators in the SO(5)
Lie algebra change the eigenvalues of $H_{1}$ and $H_{2}$.The central dots
in the root patterns are doubly degenerate. }
\label{irrep}
\end{figure}

In the following, we study the two irreducible representations given by Eqs.(%
\ref{vec}) and (\ref{sp}). First we consider the vector representation, Eq.(%
\ref{vec}). It is straightforward to show the energy spectrum of $H_{1}$ is $%
E=(-\Delta _{1},-\Delta _{2},0,\Delta _{2},\Delta _{1})$ where $\Delta
_{1,2}=\sqrt{A\pm \sqrt{A^{2}+B-C}}$ and
\begin{eqnarray}
A &=&\sum_{a<b}m_{ab}^{2}/2=\Delta ^{2}/2,  \notag \\
B
&=&2%
\sum_{a<b<c<d}(m_{ac}m_{ad}m_{bc}m_{bd}-m_{ab}m_{ad}m_{bc}m_{cd}+m_{ab}m_{ac}m_{bd}m_{cd}),
\notag \\
C &=&\sum_{a<b}\sum_{a<c<d}m_{ab}^{2}m_{cd}^{2}(1-\delta _{bc})(1-\delta
_{bd}).
\end{eqnarray}%
The single site ground state energy reaches the minimum when
$B=C$, where the energy spectrum of $H_{1}$ is $E=\{-\Delta
,0,0,0,\Delta \}$. Now we ask how is the SO(5) symmetry broken
under such a condition. Let us use $H_{1}$ as one of the two
Cartan generators\cite{georgi}. For the other Cartan generators
$H_{2}$, we choose it to be a different linear combination
of $L_{ab}$ so that  $\mathrm{Tr}%
(H_{1}H_{2})=0$. The root and weight diagrams are shown in Fig.~(\ref{irrep}%
a). Here the x and y coordinates of the dots are the eigenvalues
of $H_{1}$ and $H_{2}$, respectively. The arrows in the root
diagrams indicate how the rasing/lowering operators\cite{georgi}
in the SO(5) Lie algebra change the eigenvalues of $H_{1,2}$.
$H_{2}$ and the rasing/lowering operators generate an SO(3)
subgroup which commutes with $H_{1}$. As the result, the SO(5)
symmetry is broken down to SO(3)$\times $SO(2), where the SO(2) is
generated by $H_{1}$ itself. \newline

Second we consider the spinor representation, Eq.(\ref{sp}). It is
straightforward to show that the energy spectrum of $H_{1}$ is
$E=(-\Delta _{1},-\Delta _{2},\Delta _{2},\Delta _{1})$ with
$\Delta _{1,2}=\sqrt{A\pm \sqrt{C-B}}/\sqrt{2}$. In this case the
single site ground state energy reaches the minimum when
$A=\sqrt{C-B}$, where the energy spectrum of $H_{1}$ is
$E=\{-\Delta ,0,0,\Delta \}$. The root and weight patterns are
shown in
Fig.~(\ref{irrep})(b). Again the SO(5) symmetry is broken down to SO(3)$%
\times $SO(2).

After fixing the ratio of $m_{ab}$, we assume that the low energy
fluctuations correspond to smooth space-time dependent SO(5) rotations of
such an order parameter pattern. The NL$\sigma $ model precisely describes
such smooth fluctuations. The order parameter lives on the manifold  $\frac{SO(5)}{%
SO(3)\times SO(2)}$, which has $\mathbb{Z}$ as its second homotopy group\cite{Hatcher}. Thus the
corresponding NL${\sigma }$ model may contain a topological term, which can
lead to a spectral difference between the vector and spinor representations.

\section{The single-site Berry's phase}

To study the possible topological term, we begin by analyzing the Berry's
phase of a single SO(5) spin described by the following time-dependent
Hamiltonian
\begin{equation}
H_{1}(t)=-\sum_{a<b}m_{ab}(t)L_{ab},
\end{equation}%
where $m_{ab}(t)$ satisfy the constraints: (a) $\sum_{a<b}m_{ab}^{2}=1$, and
(b) $H_{1}(t)$ possesses SO(3)$\times $SO(2) symmetry. Both constraints can
be satisfied by starting with a reference Hamiltonian $H_{1,0}$ satisfying
(a) and (b) and perform time-dependent SO(5) conjugation, i.e.,
\begin{equation}
H_{1}(t)=U^{\dagger }(t)H_{1,0}U(t).  \label{2s}
\end{equation}
%For the vector representation an example of $H_{1,0}$ is $L_{12}$, and for
%the spinor representation an example is $\left(L_{12}-L_{34}\right)/\sqrt{2}$.

As usual, the Berry's phase is given by the loop integral of the Berry
connection\cite{wilczek}. We can use Stoke's theorem to convert this loop
integral to an areal integral over a disk with the loop as the boundary. The
advantage of doing so is the Berry curvature rather than the Berry
connection appears in the latter integral. This makes the integral involving only local quantities and  gauge invariant:
\begin{equation}
S_{B}=\frac{i}{2}\int_{0}^{1}du\int dt~\epsilon ^{\mu \nu }\text{Tr}F_{\mu
\nu }.  \label{sb}
\end{equation}%
In the above $F_{\mu \nu }=\left( \partial _{\mu }A_{\nu }-\partial _{\nu
}A_{\mu }\right) $ and $A_{\mu }=-i\langle \Omega |\partial _{\mu }|\Omega \>
$. Here $|\Omega (t,u)\rangle $ is the ground state of
\begin{equation}
H_{1}(t,u)=U^{\dagger }(t,u)H_{1,0}U(t,u).  \label{1s}
\end{equation}%
In Eq.~(\ref{1s}), $U(t,u)$ is the extension of the $U(t)$ in Eq.~(\ref{2s})
to the disk. Because $\pi _{1}$(SO(5)/SO(3)$\times $SO(2))=0, we can always
construct the extension so that $U(u=1,t)=U(t)$ and $U(u=0,t)=U_{0}$, where $%
U_{0}$ is a certain reference SO(5) element.

Using the first order perturbation theory for wave functions, it is simple
to show that
\begin{equation}
\text{Tr}F_{\mu \nu }=-i\sum_{k}\frac{\left\langle \Omega \left| \partial
_{\mu }H\right| k\right\rangle \left\langle k\left| \partial _{\nu }H\right|
\Omega \right\rangle -(\mu \leftrightarrow \nu )}{\left( E_{0}-E_{k}\right)
^{2}}.  \label{f1}
\end{equation}%
Here $k$ labels the excited states. Since all Hamiltonians described by Eq.~(%
\ref{1s}) are unitary conjugate of one another, they have the same
eigenspectrum. Under that condition we have
\begin{equation}
\partial _{\mu }H=\sum_{k}(E_{k}-E_{0})(|\partial _{\mu }k\>\langle
k|+|k\>\langle \partial _{\mu }k|).  \label{pmh}
\end{equation}

In writing down the above equation we have made a shift of the zero of
energy so that $E_0\to 0$. Substitute Eq.~(\ref{pmh}) into Eq.~(\ref{f1})
and use the fact that $\langle \Omega|k\rangle=\langle k|\Omega\rangle=0$ we
find
\begin{equation}
\mathrm{Tr} F_{\mu\nu} =-i\mathrm{Tr}(Q[\partial_\mu Q,\partial_\nu Q]),
\label{f2}
\end{equation}
where $Q$ is the ground state projection operator
\begin{equation}
Q(t,u)=|\Omega \>\langle \Omega |=U(t,u)PU^{\dagger }(t,u),  \label{q}
\end{equation}%
where $P=|0\>\langle 0|$ is the ground state projector operator of $H_{1,0}$%
. Substituting Eq.~(\ref{f2}) into Eq.~(\ref{sb}) we obtain
\begin{equation}
S_{B}=\int_{0}^{1}du\int dt\text{Tr} (Q[\partial _{u }Q,\partial _{t}Q]) .
\label{sb1}
\end{equation}

Eq.~(\ref{sb1}) actually applies for any target space. For
example, in the case of SO(3)/SO(2), we have
\begin{equation*}
U(t,u)=\left(
\begin{array}{cc}
z_{1} & z_{2} \\
-\bar{z}_{2} & \bar{z}_{1}%
\end{array}%
\right) \mathrm{~}\text{and}\mathrm{~}P=\left(
\begin{array}{cc}
1 & 0 \\
0 & 0%
\end{array}%
\right) ,
\end{equation*}%
where $z_{1,2}(t,u)$ satisfy
\begin{equation*}
\left( \bar{z}_{1},\bar{z}_{2}\right) \cdot \vec{{\sigma }}\cdot \left(
\begin{array}{c}
z_{1} \\
z_{2}%
\end{array}%
\right) =\hat{n}(t,u).
\end{equation*}%
Substitute the above two expressions into Eq.~(\ref{q}) and Eq.~(\ref{sb1})
we obtain
\begin{equation}
S_{B}=\frac{i}{2}\int_{0}^{1}du\int dt\left( \hat{n}\cdot \partial _{u}\hat{n%
}\times \partial _{t}\hat{n}\right) ,
\end{equation}%
which is the well known expression for the Berry's phase of a spin-1/2\cite%
{haldane}.

Because the dimension of our target space is eight, there are more
than one disks having the loop in question as the boundary.
Therefore, we need to ask whether Eq.~(\ref{sb1}) yields the same
answer for the Berry phase when different extensions of
$U(t)\rightarrow U(t,u)$ are used. The difference in the Berry
phase using two different disks as extension can be calculated by
integrating the Berry curvature over a closed two dimensional
surface formed by joining the two disks at their common boundary.
The resulting closed surface has the topology of a 2-sphere.
Because the second homotopy group of our target space is
$\mathbb{Z}$, all 2-spheres in the target space are topologically
the multiple of a basic sphere. Hence all we need to check is
whether the Berry curvature integral is integer multiple of $2\pi
$ when $t$ and $u$ in Eq.~(\ref{sb1}) parameterize the basic
2-sphere\cite{witten}. In the following, we perform such a
calculation.

For the vector representation, we choose $H_{1,0}=L_{12}$, and pick $U(t,u)$
so that
\begin{equation}
U^{\dagger }(t,u)H_{1,0}U(t,u)=\hat{w}(t,u)\cdot \vec{L},  \label{ex}
\end{equation}%
where $\vec{L}=(L_{13},L_{23},L_{12})$,
\begin{equation*}
\hat{w}_{i}(t,u)=(\sin (\pi u)\cos \frac{2\pi t}{\beta },\sin (\pi u)\sin
\frac{2\pi t}{\beta },\cos (\pi u)),
\end{equation*}
and $\beta $ is the period of the imaginary time. For the spinor
representation, we take $H_{1,0}=(L_{12}-L_{34})/\sqrt{2}$, and $\vec{L}%
=\left( L_{13}+L_{24},L_{14}-L_{23},L_{12}-L_{34}\right) /\sqrt{2}$.

Using Eq.~(\ref{sb1}), or equivalently Eqs.(\ref{sb}) and (\ref{f1}), we
find that
\begin{equation}
S_{B,\mathrm{basic~sphere}}^{\mathrm{~vector}}=4\pi ,\text{ \ \ }S_{B,%
\mathrm{basic~sphere}}^{\mathrm{~spinor}}=2\pi .  \label{dif}
\end{equation}%
Thus the condition for the uniqueness of the Berry phase is satisfied. As we
shall see, the difference between the vector and spinor Berry phase in Eq.~(%
\ref{dif}) serves to distinguish the vector and spinor SO(5) chains.

\section{The lattice Berry's phase, gapful versus gapless, and the edge
states}

We now extend the above single-site Berry phase analysis to the one
dimensional lattice. If the order parameter is perfectly
``antiferromagnetic'' at all time, the ground state of $%
H_{1}(t)$ are $|\Omega \rangle (t)$ on the even sublattice and its conjugate
$|\bar{\Omega}(t)\>=R|\Omega (t)\>^{\ast }$ on the odd sublattice, where $R$
is the operator that satisfies $RL_{ab}^{\ast }R^{-1}=-L_{ab}$. For the
vector representation, $R$ is given by $%
R_{15}=-R_{24}=R_{33}=-R_{42}=R_{51}=-1$ and $R_{ij}=0$ otherwise. For the
spinor representation, $R=-iI_{2\times 2}\otimes \sigma _{y}$. Using the
above result, it is straightforward to show that $\langle \Omega
|L_{ab}|\Omega \>=-\langle \bar{\Omega}|L_{ab}|\bar{\Omega}\>$, and $%
RU^{\ast }R^{-1}=U$ for all $U\in $ SO(5). This, plus the invariance of the
trace under matrix transposition, allows one to show that $\epsilon ^{\mu
\nu }\text{Tr}\bar{Q}\partial _{\mu }\bar{Q}\partial _{\nu }\bar{Q}%
=-\epsilon ^{\mu \nu }\text{Tr}Q\partial _{\mu }Q\partial _{\nu }Q$. As the
result, the Berry's phases associated with neighboring sites tend to cancel
each other. Let $r$ label the center of mass position of site $i$ and $i+1$
for $i=$ odd, the total lattice Berry's phase is equal to
\begin{equation}
S_{B}^{\mathrm{tot}}=\sum_{r}\sum_{\epsilon =\pm 1}(-1)^{\epsilon
-1}\int_{0}^{1}du\int dt\mathrm{Tr}(Q_{r+\epsilon /2}[\partial
_{u}Q_{r+\epsilon /2},\partial _{t}Q_{r+\epsilon /2}]),  \label{gk}
\end{equation}%
where $Q$ is a smooth function of spacial coordinates. Under such a
condition, Eq.~(\ref{gk}) has a continuum limit
\begin{equation}
S_{B}^{\mathrm{tot}}=\frac{1}{2}\int dx\int dt\text{Tr}Q[\partial
_{t}Q,\partial _{x}Q],  \label{sb3}
\end{equation}%
the factor of 1/2 arises from the density of odd lattice sites.
Under open boundary condition, Eq.~(\ref{sb3}) becomes
\begin{eqnarray}
S_{B}^{\mathrm{tot}} &=&{\frac{1}{2}}\int dxdt\mathrm{Tr}Q[\partial
_{t},Q\partial _{x}Q]  \notag \\
&&+{\frac{1}{2}}\int_{0}^{1}du\int dt\left\{ \mathrm{Tr}[Q[\partial
_{u},Q\partial _{t}Q]]_{R}-\mathrm{Tr}[Q[\partial _{u}Q,\partial
_{t}Q]]_{L}\right\} ,  \label{tbe}
\end{eqnarray}%
where the subscript ``R''\ and ``L''\ labels the right and the left ends.
This topological term together with the stiffness term from the energetics,
constitute the NL${\sigma }$ model for the SO(5) spin chain.

Eqs.~(\ref{dif}) and (\ref{sb3}) have important implications. The fact that
the mapping from the space-time to the target space is classified by integer
homotopy classes implies that the space-time order parameter configurations
can be grouped into different topological sectors labeled by an integer
topological invariant. This is similar to the SO(3) NL${\sigma }$ model
where the topological invariant, the Pontryagin index\cite{pont}, is the
number of times which the order parameter configurations cover the target
space $S^{2}$. In our case there is an analogous integer topological index,
which we will refer to as the Pontryagin index as well. Eq.~(\ref{dif})
implies that for the vector SO(5) spin chain the Berry phase associated with
the order parameter configuration having different Pontryagin indices are
all the same because $\exp (i4\pi /2\times \mathrm{integer})=+1$. Given the
facts that (i) the topological term has no effect (hence the NL$\sigma $
model has only the stiffness terms), and (ii) the target space dimension is
high, it is easy to believe that the vector SO(5) spin chain should have a
quantum disordered, i.e., translational invariant gapped, phase. For the
spinor SO(5) chain, however, the order parameter configurations with even
Pontryagin index have the Berry phase $\exp (i2\pi /2\times \mathrm{%
even~integer})=+1$, while those with odd Pontryagin index have Berry phase $%
\exp (i2\pi /2\times \mathrm{odd~integer})=-1$. This result is exactly
analogous to the Berry's phase in the spin-1/2 representation of the SO(3)
antiferromagnetic Heisenberg chain. In view of the result of Ref.\cite{xc},
we conclude that the above non-trivial Berry's phase also implies the lack
of an energy gap as long as the translation symmetry is unbroken.

Now we comment on the edge state of the SO(5) ``valence bond solid state''
in Ref.\cite{zx}. By tuning the ratio of $J_{1}$ and $J_{2}$ in Eq.(\ref{h5}%
), Tu et al were able to show that a short-range entangled, translational
invariant matrix product state is the exact ground state. In addition, under
the open boundary condition the ground state wavefunction becomes $4\times
4=16$ fold degeneracy. According to Eq.~(\ref{tbe}), the boundary of a vector
spin chain should exhibit the following Berry phase
\begin{equation}
{\frac{1}{2}}\int_{0}^{1}du\int dt\mathrm{Tr}Q[\partial _{u},Q\partial
_{t}Q].  \label{sf}
\end{equation}%
When $Q(t,u)$ is a unit Pontryagin index order parameter
configuration, the value of Eq.(\ref{sf}) is ${\frac{1}{2}}*{4\pi
}=2\pi $. This is consistent with the spinor Berry phase in
Eq.~(\ref{dif}). Therefore, the edge state of the vector SO(5)
spin chain carries the spinor representation. Because the latter
is 4-dimensional, each end of the chain independently yields a
4-fold degeneracy of the ground state, resulting in a total
$4\times 4=16$ fold degenerate ground state for the open chain.

Before closing a technical remark is in order. The readers might wonder what
if the ratio between different components of $m_{ab}$ fluctuates away from
the the optimal value. When that happens the single site spectrum will
become $\{-\Delta _{1},-\Delta _{2},0,\Delta _{2},\Delta _{1}\}$ and $%
\{-\Delta _{1},-\Delta _{2},\Delta _{2},\Delta _{1}\}$ for the
vector and spinor representation, respectively. In this case, the
SO(5) symmetry is broken down to SO(2)$\times $SO(2). The second
homotopy group of $\frac{SO(5)}{SO(2)\times SO(2)}$ is
$\mathbb{Z}\oplus \mathbb{Z}$ rather than $\mathbb{Z}$. In other
words, the image of the space-time in the target space is
topologically the multiple of two basic spheres. As $\Delta
_{2}\rightarrow 0$, one of these spheres shrinks to a point. We
have checked that so long as $\Delta _{2}$ is small, i.e., when
the ratio between $m_{ab}$ does not deviate from the optimal value
too drastically, the Berry phase is only sensitive to the
Pontryagin index of the dominant (large) sphere. Hence all the
results discussed earlier remain unchanged.

\section{Conclusion}

We have studied the Berry's phase of the antiferromagnetic SO(5)
spin chain, and shown the existence of a topological term in the
non-linear $\sigma$ model description of the system. The quantum
phase factor associated with this topological term differentiates
the vector (linear) and spinor (projective) representations. We
argue that this leads to the spectral difference as long as the
translation and SO(5) symmetry is unbroken. More specifically, the
vector spin chain can have a totally symmetric ground state while
having an energy gap. The spinor chain, on the other hand, must be
gapless if there is no symmetry breaking. Under the open boundary
condition, we find the boundary Berry's phase of the vector spin
chain is consistent with the derived edge degeneracy of an exactly
solvable model. The present result can be straightforwardly
generalized to other irreducible representations, leading to two
classes of SO(5) spin chain: in one class the site representation
is linear, and in the other the site representation is projective.
While the first class can have a totally symmetric ground state
and maintain a gapped spectrum, the second class must have
gapless spectrum if there is no symmetry breaking. %This result generalizes
%Haldane's seminal works\cite{haldane} to a higher rank Lie group.

Finally, we would like to make contact with relevant works in the
literature. Eq.(\ref{sb1}) appeared in Ref.\cite{wiegmann} in
discussing a super-symmetric version of the Hubbard model. Of
course, the target space is entirely different. In Ref.\cite{aff}
Affleck discussed the topological term of an SU(N) spin chains. In
that case the target space is $U(N)/U(m)\times U(N-m)$, which also
has $\mathbb{Z}$ as its second homotopy group. In
Ref.\cite{read1}, Read and Sachdev have discussed the effects of
Haldane's (SU(2) or SO(3)) Berry phase term in two space
dimensions\cite{read1}. Subsequently, they generalized this to the
SU(N) group in Ref.\cite{read}. The essential difference between
the SU(N) and the SO(5) spin chains is their target spaces.
Different target spaces have distinct topology. As far as we know,
this is the first time one points out that the target space of the
SO(5) spin chain also has second homotopy group $\mathbb{Z}$, and
constructs the Berry's phase term for the NL$\sigma$ model
description. Of course, analogous to Refs.\cite{read1,read} one
can ask what is the effect of this topological term in two space
dimensions. We leave this question for future studies.

\section{Acknowledgement}

We thank Wu-Yi Hsiang, Xiao-Gang Wen, Zheng-Cheng Gu, Ari Turner, Xie Chen,
and Geoffrey Lee for helpful discussions. DHL is supported by DOE grant
number DE-AC02-05CH11231 and thanks for the hospitality for the MIT
condensed matter theory group in the last stage of this work. GMZ and TX
acknowledge the support of NSF-China and the National Program for Basic
Research of MOST, China.

\end{document}